\documentclass[twocolumn]{article}
\usepackage{multicol}
\usepackage[dvips]{graphicx}
\usepackage{amsmath}
\usepackage{epsfig}
\usepackage{booktabs}
\usepackage{fancyhdr}

\newcommand{\bfr}{\begin{flushright}}
\newcommand{\efr}{\end{flushright}}
 
\begin{document}
\twocolumn[
\title{New Solutions of the $2+1$ Dimensional Einstein Gravity Coupled
to Maxwell Power type  Non Linear Electric field with Dilaton field 
}
\author{Masashi Kuniyasu\thanks{u003wa@yamaguchi-u.ac.jp}
\\
%\address
{\small
Graduate School of Science and Technology, Yamaguchi University,}\\
{\small Yamaguchi-shi, Yamaguchi 753--8512, Japan}
}
\date{\today
}

\maketitle
\begin{abstract}
New solutions are derived in the $2+1$ gravity which is coupled to
$|{\cal F}|^k$ type non-linear electric field in Maxwell
Power theory with dilaton field. We obtain consistent solutions in
general $k$ case. We also investigate the behavior of
the metric function with the space-time singularity.
Then, we found some black hole solutions when the space-time has
a singular point at $r=0$. 
\\
~
\\
%PACS number(s): ????
\end{abstract}]
%\PACS{}

\thispagestyle{fancy}
\lfoot{$^*$u003wa@yamaguchi-u.ac.jp}
\renewcommand{\headrulewidth}{0pt}
\renewcommand{\footrulewidth}{0.4pt}
%*****************************************************************
\section{Introduction}
%*****************************************************************
Recently, one of the most standard theory of gravitation is
the Einstein's general theory of relativity (GR).
However, GR is the classical theory, so we have to quantize the
theory. To construct the quantum GR, many modified theories of
gravitation had suggested. As there is a Schwarzschild black hole
(BH) solution in the GR, modified gravity also have
characteristic solutions with theory to theory. That is why to
find a new solution in the theory of gravitation is important.

The $2+1$ gravity has been suggested by
Deser and Jackiw~\cite{SR}. This theory has the characteristic
structure as follows. At first, lower dimensional theory of
gravitation has fewer unknown functions than higher dimensional
theory. That is why to derive new exact solutions in $2+1$
space-time is more easier than higher dimensional one.
Next, Weyl tensor always vanishes in the gravity theory which
has only $3$ or fewer dimensional space-time. In addition, the
lack of degrees of freedom tends to gravitational wave does not
propagate. Carlip reviewed such as interesting properties of
the $2+1$ gravity~\cite{SC}.

This theory came to attract attention after the discovery of
BTZ solution~\cite{BTZ}.
A solution represents the black hole which is in the AdS
space-time. Note that BTZ black hole is not asymptotically
flat like Schwarzschild black hole. After their research, a lot of
solutions were derived with many types of matter fields in the
$2+1$ space-time.

In the present article, let us treat non linear electric field (NLEF)
and dilaton field as matters. To the past research, Cataldo find exact
solutions with Born-Infeld~\cite{BI} type and $| {\cal F}^{3/4} |$
type NLEF \cite{MA}~,~\cite{CA}. Note that ${\cal F}$ is the
Maxwell invariant which is defined as ${\cal F}=F_{\mu\nu}F^{\mu\nu}$
where $F_{\mu\nu}=\partial_\mu A_\nu-\partial_\nu A_\mu$ is the
electric tensor in which $A_\mu$ is vector potential. Gurtug found
an exact solution with $| {\cal F}^k |$ ($k$ is a real number) type 
NLEF which is called Maxwell Power (MP) theory~\cite{GHH}. They are
circular symmetric static solution with electric field.
Moreover, solutions in the general dimension are derived by some
authors~\cite{HM07},~\cite{HM08},~\cite{MHM}.
Then, Hendi found that MP type NLEF is related to the spherically
symmetric solution of the $F(R)=R^k$ type modified gravity~\cite{SH}.
Recently, new solutions were derived with scalar hair in the AdS
space-time~\cite{WD}.

To treat the dilaton field is meaning full because it is related
to the open and the closed bosonic string theory (on our paper,
we however neglect the Kolb-Ramond antisymmetric tensor field).
Dilaton field is also important for the theory of gravitation
because it has correspondence to the modified gravity.
Since to obtain solutions of the theory is important, researchers
find solutions in the theory of gravitation with dilaton field as
follows. Gibbons and Maeda found a electrically charged black hole
solution in $3+1$ space-time~\cite{GM}. In the $2+1$ gravity, S\'a
and Lemos found the black hole solution without electric field~\cite{PJ}.
Then,  Chan and Mann discovered some solutions with linear electric
field~\cite{KR}. After their research, Yamazaki and Ida introduced
Born-Infeld type NLEF and found a new solution~\cite{RI}.
However, MP type NLEF with dilaton field had not considered.
That is why we try to find new solutions of the $2+1$ gravity
which is coupled to MP type NLEF with dilaton field.

In this article, we take natural unit which is $c=8\pi G=1$, where
$c$ is speed of light, $G$ is Newton constant in the $2+1$
space-time. Additionally, we set the metric signature as $(-++)$.

%%%%%%%%%%%%%%%%%%%%%%%%%%%%%%%%%%%%%%%%%%%%%%%%%%%%%%%%%%%%%%%%%%
\section{Circular symmetric static solutions of the $2+1$ gravity
which is coupled to ${\cal F}^k$ type NLEF with dilaton field}
%%%%%%%%%%%%%%%%%%%%%%%%%%%%%%%%%%%%%%%%%%%%%%%%%%%%%%%%%%%%%%%%%% 
%%%%%%%%%%%%%%%%%%%%%%%%%%%%%%%%%%%%%%%%%%%%%%%%%%%%%%%%%%%%%%%%%%
\subsection{Action and basic field equations }
%%%%%%%%%%%%%%%%%%%%%%%%%%%%%%%%%%%%%%%%%%%%%%%%%%%%%%%%%%%%%%%%%%
The action of the $2+1$ gravity which is coupled to ${\cal F}^k$
type NLEF with dilaton field is
\begin{equation}
\begin{split}
I=\frac{1}{2}
&\int d^3x\sqrt{-g} \\
&\times\left\{R-\alpha\left(\partial \phi
 \right)^2-e^{-\phi}\left| {\cal F} \right|^k-2e^{\beta\phi}
 \Lambda \right\} \,.
\label{eq2-1}
\end{split}
\end{equation}
Where $g$ is determinant of the metric $g_{\mu\nu}$, $R$ is
Ricci scalar, $\phi$ is the dilaton field, $\alpha$ and
$\beta$ are constants. We get Einstein equation  when we take the
variation of $g_{\mu\nu}$,
\begin{equation}
G_\nu^\mu+\Lambda e^{\beta\phi}\delta_\nu^\mu=T_\nu^\mu \,,
\label{eq2-2}
\end{equation}
where $T^\mu_\nu$ is energy-momentum tensor
\begin{equation}
\begin{split}
T^\mu_\nu=
&\alpha\partial^\mu\phi\partial_\nu\phi-\frac{\alpha}
 {2}\left(\partial\phi\right)^2\delta^\mu_\nu \\
&-\left|{\cal F}\right|^k\frac{e^{-\phi}}{2}\left(\delta^\mu_\nu
-\frac{4kF^{\mu\nu}F_{\mu\nu}}{{\cal F}}\right)\,.
\label{eq2-3}
\end{split}
\end{equation}
On the other hand, let us take the variation of (\ref{eq2-1})
respect to $\phi$, we get the following dilaton equation
\begin{equation}
\alpha\nabla_\mu\left(\partial^\mu\phi\right)+\frac{
e^{-\phi}}{2}\left|{\cal F}\right|^k
-\beta e^{\beta\phi}\Lambda =0
\label{eq2-4} \,,
\end{equation}
In addition, electric field equation is derived from the variation
of (\ref{eq2-1}) which is respect to $A_\mu$
\begin{equation}
\partial_\mu\left( \sqrt{-g}e^{-\phi}F^{\mu\nu}
\left|{\cal F}\right|^{k-1}\right)=0 \,.
\label{eq2-5}
\end{equation}

%%%%%%%%%%%%%%%%%%%%%%%%%%%%%%%%%%%%%%%%%%%%%%%%%%%%%%%%%%%%%%%%%%
\subsection{Static circular symmetric solutions}
%%%%%%%%%%%%%%%%%%%%%%%%%%%%%%%%%%%%%%%%%%%%%%%%%%%%%%%%%%%%%%%%%%
Now, let us solve above equations under the circular symmetric
static ansatz. We set metric ansatz as
\begin{equation}
ds^2=-f(r)^2dt^2+\frac{e^{2\delta (r)}dr^2}{f(r)^2}
+r^2d\theta ^2 \,,
\label{eq2-6}
\end{equation}
where $f(r)~,~\delta(r)$ are unknown functions which is depended
only radial coordinate $r$. Following discussion, we assume
dilaton field is depended only $r$. Beside, electric field ansatz
is
\begin{equation}
{\bf F}=e^{\phi (r)+\delta (r)}E(r)dt\wedge dr \,,
\label{eq2-7}
\end{equation}
where $E(r)$ and $\phi(r)$ are unkown functions which is
depended only $r$.

Then, Einstein equations will be
\begin{equation}
\begin{split}
&\frac{e^{-2\delta (r)}\left(f^2\delta_r-ff_r\right)}{r} \\
&=\frac{\alpha}{2}e^{-2\delta}f^2\left(\frac{d\phi}{dr}\right)^2
+\frac{1-2k}{2}e^{-\phi}\left|{\cal F}\right|^k+e^{\beta\phi}
\Lambda\,,
\label{eq2-8}
\end{split}
\end{equation}
\begin{equation}
\begin{split}
\frac{e^{-2\delta (r)}ff_r}{r}
&=\frac{\alpha}{2}e^{-2\delta}f^2\left(\frac{d\phi}
 {dr}\right)^2 \\
&-\frac{1-2k}{2}e^{-\phi}\left|{\cal F}\right|^k-e^{\beta\phi}
\Lambda\,,
\label{eq2-9}
\end{split}
\end{equation}
\begin{equation}
\begin{split}
&e^{-2\delta (r)}\left(ff_r
 \delta_r-ff_{rr}-(f_r)^2\right) \\
&=\frac{\alpha}{2}e^{-2\delta}f^2\left(
\frac{d\phi}{dr}\right)^2-\frac{1}{2}e^{-\phi}
\left|{\cal F}\right|^k+e^{\beta\phi}\Lambda\,.
\label{eq2-10}
\end{split}
\end{equation}
When we subtract (\ref{eq2-9}) from (\ref{eq2-8}), we get
\begin{equation}
\frac{1}{r}\frac{d\delta}{dr}=\alpha\left(\frac{d\phi}
{dr}\right)^2\,.
\label{eq2-11}
\end{equation}
In this paper, we assume function $\delta(r)$ as
\begin{equation}
\delta(r)=n\ln\left(\frac{r}{r_*}\right)\,,
\label{eq2-12}
\end{equation}
which is embraced in article~\cite{RI}. Here $n$ and $r_*$ are
constants. Thus, we can determinant the dilaton field $\phi$ as
\begin{equation}
\phi(r)=\pm\sqrt{\frac{n}{\alpha}}\ln\left(\frac{r}{r_0}\right)\,,
\label{eq2-13}
\end{equation}
where $r_0$ is some constant.

Next, let us discuss the electric field equation. We can treat
equation (\ref{eq2-5}) analytically, then
\begin{equation}
E^{2k-1}(r)=\frac{q}{r}\left(\frac{r_0}{r}\right)^{\pm2\sqrt{
n/\alpha}(k-1)}\,,
\label{eq2-14}
\end{equation}
where $q$ is an integration constant which is related to the charge of
the space-time. The $k=1$ case, when the electric field $E(r)$
proportional to $\ln(r)$, was already discussed by Chan~\cite{KR}.

To determine the relation of constants, we calculate Einstein
equations and dilaton field. At first, take the summation
(\ref{eq2-8}) and (\ref{eq2-9}), then
\begin{equation}
\frac{e^{-2\delta}(f^2\delta_r-2ff_r)}{r}
=(1-2k)e^{-\phi}\left|{\cal F}\right|^k+2e^{\beta\phi}\Lambda\,.
\label{eq2-15}
\end{equation}
On the other hand, from the dilaton field equation, we get
\begin{equation}
\begin{split}
&\alpha e^{-2\delta}\left\{\frac{f^2}{r}-f^2\delta_r+2ff_r
 \right\}\left(\frac{d\phi}{dr}\right) \\
&+\alpha e^{-2\delta}f^2\frac{d^2\phi}{dr^2}+\frac{e^{-\phi}}{2}
\left|{\cal F}\right|^k-\beta e^{\beta\phi}\Lambda=0\,.
\label{eq2-16}
\end{split}
\end{equation}
Thus we get the following relation when we use equation
(\ref{eq2-12}),  (\ref{eq2-13}) and (\ref{eq2-15}), (\ref{eq2-16})
\begin{equation}
\begin{split}
&\left\{\pm\sqrt{n\alpha}(1-2k)
 -\frac{1}{2}\right\}e^{-\phi}|{\cal F}|^k \\ 
&+\left(\pm2\sqrt{n\alpha}+\beta\right)e^{\beta\phi}\Lambda=0\,.
\label{eq2-17}
\end{split}
\end{equation}
This equation implies that there are two types of the solution
in our discussion. We call the ``first class'' solution which
means the coefficients $e^{-\phi}|{\cal F}|^k$ and
$e^{\beta\phi}$ vanishes at non zero $\Lambda$;
\begin{equation}
-\beta=\pm2\sqrt{n\alpha}\,,
\label{eq2-18}
\end{equation}
\begin{equation}
\beta\left(1-2k\right)+1=0\,.
\label{eq2-19}
\end{equation}
This class of the solutions are not consistent to $k=1/2$
type MP theory because they are not consistent to the
coefficient of $e^{-\phi}|{\cal F}|^k$ equal to zero.

Other types of the solution which we call ``second class''
solutions mean
\begin{equation}
e^{-\phi}|{\cal F}|^k  
=-\frac{\pm2\sqrt{n\alpha}+\beta}
{\pm\sqrt{n\alpha}(1-2k)-1/2}\Lambda e^{\beta\phi}\,.
\label{eq2-17a}
\end{equation}
This means that coefficients do not vanish. It also means
we could get the consistent solution at $k=1/2$ case.
Hereafter, we will get the solutions in the explicit form.

%%%%%%%%%%%%%%%%%%%%%%%%%%%%%%%%%%%%%%%%%%%%%%%%%%%%%%%%%%%%
\subsubsection{The first class solutions}
%%%%%%%%%%%%%%%%%%%%%%%%%%%%%%%%%%%%%%%%%%%%%%%%%%%%%%%%%%%%
We now consider the first class solutions. In this case, we
get the following differential equation
\begin{equation}
\begin{split}
&\frac{2f}{r}\frac{df}{dr}-n\frac{f^2}{r^2} \\
&=(2k-1)\left(\frac{r}{r_*}\right)^{2n}\left(\frac{r}
 {r_0}\right)^{8nk^2(1-k)/(2k-1)} \\
&\times\left(\frac{q}{r}\right)^{2k/(2k-1)}
 -2\left(\frac{r_0}{r_*}\right)^{2n}\Lambda\,.
\label{eq2-20}
\end{split}
\end{equation}
This equation can solve analytically, then
\begin{equation}
\begin{split}
f(r)
&=-\left(\frac{r_0}{r_*}\right)^{2n}\frac{2\Lambda r^2}{2-n}-mr^n \\
&+\left(\frac{r_0}{r_*}\right)^{2n}\frac{r^2(2k-1)^2}{2(k-1)(1-4n
 k^2)-n(2k-1)} \\
&\times\left(\frac{r}{r_0}\right)^{8nk^2(1-k)/(2k-1)}
 \left(\frac{q}{r}\right)^{2k/(2k-1)}\,,
\label{eq2-21}
\end{split}
\end{equation}
where $m$ is an constant. Again, this solution is not consistent to
the $k=1/2$ MP theory because of the constrain (\ref{eq2-19}).

%%%%%%%%%%%%%%%%%%%%%%%%%%%%%%%%%%%%%%%%%%%%%%%%%%%%%%%%%%%%%%%%%%
\subsubsection{The second class solutions}
%%%%%%%%%%%%%%%%%%%%%%%%%%%%%%%%%%%%%%%%%%%%%%%%%%%%%%%%%%%%%%%%%%
In the second class solution, the electric part
$e^{-\phi}|{\cal F}|^k$ and cosmological part $e^{\beta\phi}$ have
the same $r$ dependence. This condition restricts constants as
\begin{equation}
\pm\sqrt{\frac{n}{\alpha}}=\frac{2k}{1+\beta(1-2k)}\,.
\label{eq2-22}
\end{equation}
Then, equation (\ref{eq2-17a}) becomes
\begin{equation}
\begin{split}
E^{2k}
&=-\frac{4k\alpha+\beta(1+\beta(1-2k))}
 {4k\alpha(1-2k)-\beta(1-2k)-1}2\Lambda \\
&\times e^{(\beta-(2k-1))\phi}\,.
\label{eq2-23}
\end{split}
\end{equation}
After some calculation, we get the following differential equation
\begin{equation}
\begin{split}
&\frac{2f}{r}\frac{df}{dr}-n\frac{f^2}{r^2} \\
&=2\frac{\beta^2(1-2k)^2+8k\alpha(1-2k)-1}
 {4k\alpha(1-2k)-\beta(1-2k)-1}
 \Lambda e^{\beta\phi+2\delta}\,.
\label{eq2-24}
\end{split}
\end{equation}
This equation can solve analytically as
\begin{equation}
\begin{split}
f^2
&=-mr^\alpha-\Lambda r^2\frac{\beta^2(1-2k)^2+8k\alpha(1-2k)
 -1}{4k\alpha(1-2k)-\beta(1-2k)-1} \\
&\times\frac{(1+\beta(1-2k))^2}
 {2k^2\alpha+(1+\beta(1-2k))(1+\beta(1-k))} \\
&\times\left(\frac{r}{r_0}\right)^{2k\beta/(1+\beta(2k-1))}
 \left(\frac{r}{r_*}\right)^{8k^2\alpha/(1+\beta(1-2k))^2}\,,
\end{split}
\label{eq2-25}
\end{equation}
where $m$ is an constant.

The solution which belongs to second class is consistent to the
$k=1/2$ MP theory. In the $k=1/2$ theory, solutions of the field
equations are
\begin{equation}
r_0=q~,~\pm\sqrt{\frac{n}{\alpha}}=1\,,
\label{eq2-26}
\end{equation}
\begin{equation}
\phi(r)=\ln\left(\frac{r}{q}\right)\,,
\label{eq2-27}
\end{equation}
\begin{equation}
\delta(r)=\alpha\ln\left(\frac{r}{r_*}\right)\,,
\label{eq2-28}
\end{equation}
\begin{equation}
E(r)=2(2\alpha+\beta)\left(\frac{r}{q}\right)
^\beta\Lambda\,,
\label{eq2-29}
\end{equation}
\begin{equation}
f^2(r)=\frac{-2\Lambda r^2}{2+\alpha+\beta}\left(\frac{r}{r_*}
\right)^{2\alpha}\left(\frac{r}{q}\right)^\beta -mr^\alpha\,.
\label{eq2-30}
\end{equation}
These solutions are consistent to the solutions of the field
equations from the action
\begin{equation}
\begin{split}
I=\frac{1}{2}
&\int d^3x\sqrt{-g} \\
&\times\left\{R-\alpha\left(\partial \phi
 \right)^2-e^{-\phi}\sqrt{\left| {\cal F} \right|}-2e^{\beta\phi}
 \Lambda \right\} \,,
\label{eq2-31}
\end{split}
\end{equation}
under the circular symmetric ansatz (\ref{eq2-6}) and
(\ref{eq2-7}). It is characteristic structure of the second class
solutions.  In the past
work, special electric field ansatz that treated by  Mazharimousavi
\cite{MH} is also consistent to the $k=1/2$ type MP theory.
However, $k=1/2$ type MP theory has no consistent
solution without dilaton field or first class solutions with
circular symmetric ansatz which adopted in our paper. 

%%%%%%%%%%%%%%%%%%%%%%%%%%%%%%%%%%%%%%%%%%%%%%%%%%%%%%%%%%%%%%%%%%
\section{Fundamental structure of the space-time}
%%%%%%%%%%%%%%%%%%%%%%%%%%%%%%%%%%%%%%%%%%%%%%%%%%%%%%%%%%%%%%%%%%
In this chapter, let us consider the fundamental geometric
quantity such as scalar invariant. Addition, we investigate
the behavior of the metric function $f(r)$.

As was mentioned above, hereafter we will investigate negative
cosmological constant case, then
\begin{equation}
\ell=\frac{1}{\sqrt{-\Lambda}}\,,
\label{eq3-1}
\end{equation}
where $\ell$ is cosmological radius. Why we consider negative
$\Lambda$ case? At first, asymptotic behavior of the fundamental
black hole solution in the $2+1$ gravity that called BTZ black
hole is AdS space-time. Moreover, because of the AdS/CFT
correspondence~\cite{JM}, it is believed that AdS space-time
itself interesting. So, we hope that it is meaning full to include
the negative cosmological constant.

%%%%%%%%%%%%%%%%%%%%%%%%%%%%%%%%%%%%%%%%%%%%%%%%%%%%%%%%%%%%%%%%%%
\subsection{first class solutions}
%%%%%%%%%%%%%%%%%%%%%%%%%%%%%%%%%%%%%%%%%%%%%%%%%%%%%%%%%%%%%%%%%%
The solution (\ref{eq2-21})is constructed by power term of $r$.
Then, we can rewrite (\ref{eq2-21}) as
\begin{equation}
f(r)^2=Ar^2-mr^n-B\left(\frac{1}{r}\right)^c\,,
\label{eq3-2}
\end{equation}
where
\begin{equation}
\begin{split}
&A=\left(\frac{r_0}{r_*}\right)^{2n}\frac{1}
 {(2-n)\ell^2}\,, \\
&B=\frac{(2k-1)^2}{2(k-1)(1-4nk^2)-n(2k-1)}q^{2k/(2k-1)} \\
&\times r_0^{8nk^2(k-1)/(2k-1)}\,, \\
&c=\frac{2k}{2k-1}\left(1-4nk^2(1-k)\right)\,.
\label{eq3-3}
\end{split}
\end{equation}
Then, Fundamental geometric quantity such as Ricci scalar of
the solution (\ref{eq2-21}) reads to
\begin{equation}
\begin{split}
R=-\frac{6Ar^2+(c^2-5c+6)Br^{2-c}-mn(n+1)r^n}{r^2}\,.
\label{eq3-5}
\end{split}
\end{equation}
The singular behavior of the Ricci scalar is depended on $n$
and $k$ (equal to $\alpha$ and $\beta$) becomes as follows. i)~a
singular point at origin. ii)~a singular point at infinity.
iii)~two singular points at origin and infinity. Higher order
curvature invariant such as$R^{\mu\nu}R_{\mu\nu}$ and
$R^{\mu\nu\sigma\lambda}R_{\mu\nu\sigma\lambda}$ also has same
structure. Thus, our new solution (\ref{eq2-21}) has physical
singular point at $r=0$ or $r=\infty$ that depends on $n$ and $k$.
In addition, it is possible to create a black hole solution with some
fine parameter sets. To get a black hole solution, we will
choice i) type physical singular behavior. Then, parameters must
satisfy $n-2\leq0~,~C\leq0$. They are equal to
\begin{equation}
n\leq 2~,~n\leq \frac{1}{4k^2(1-k)}\,.
\label{eq3-6}
\end{equation}

\begin{figure}[htbp]
\begin{center}
\includegraphics[width=70mm]{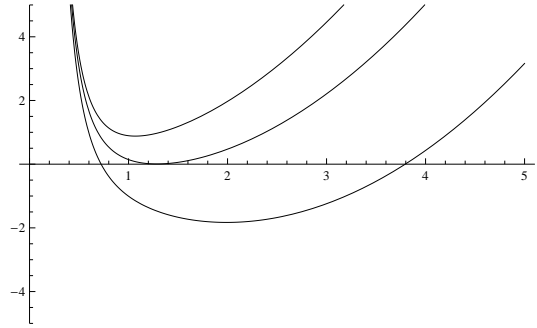}
\end{center}
\caption{The behavior of the metric function when singular point
lies at $r=0$. The horizontal line represents radial position $r$,
and vertical line represents the metric function $f(r)$. The
parameter sets are $\ell=r_0=r_*=1$ and $n=1$ with $k=3/4$. Top
line, middle line and bottom line means $m=0.1,~0.85,~2$ respectively.}
\label{fig1}
\end{figure}

\begin{figure}[htbp]
\begin{center}
\includegraphics[width=70mm]{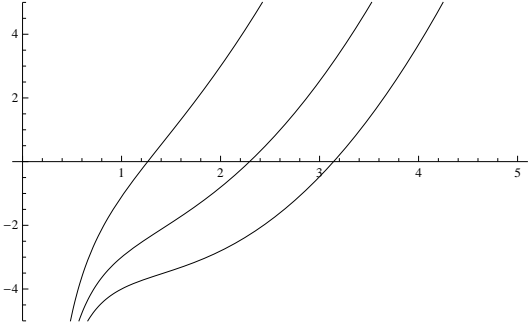}
\end{center}
\caption{The behavior of the metric function when singular point
lies at $r=0$. The parameter sets are $\ell=r_0=r_*=1$ and $n=0.1$
with $k=3/4$. Left line, middle line and right line means
$m=0.1,~2,~3$ respectively.}
\label{fig2}
\end{figure}

The behavior of the metric function in this
case is shown in Figure~\ref{fig1} and Figure~\ref{fig2}. We cannot
derive the horizon radius with analytical method in generally,
however, there are horizons at $f(r_H)=0$ where $r_H$ is its
radius at some parameter sets. From Figure~\ref{fig1}, we notice
that there are two horizons and the space-time is similar to the
R-N black hole. Thus, the parameter sets in Figure~\ref{fig1}
could generate R-N like black holes. On the other hand, some
parameter sets construct the metric function like
Figure~\ref{fig2}. Then, the space-time is similar to the
Schwarzschild-AdS black hole. That is why we conclude that our new
solutions can generate some different types of black hole.

%%%%%%%%%%%%%%%%%%%%%%%%%%%%%%%%%%%%%%%%%%%%%%%%%%%%%%%%%%%%%%%%%%
\subsection{second class solutions}
%%%%%%%%%%%%%%%%%%%%%%%%%%%%%%%%%%%%%%%%%%%%%%%%%%%%%%%%%%%%%%%%%%
Here, we consider about second class solutions as first class
solutions. Second class solutions (\ref{eq2-25}) can rewrite as
\begin{equation}
\begin{split}
f(r)^2=-mr^n+Cr^2\left(\frac{1}{r}\right)^d
\end{split}
\label{eq3-11}
\end{equation}
where
\begin{equation}
\begin{split}
&C=\frac{1}{\ell^2}\frac{\beta^2(1-2k)^2+8k\alpha(1-2k)-1}
{4k\alpha(1-2k)-\beta(1-2k)-1} \\
&\times\frac{(1+\beta(1-2k))^2}
 {2k^2\alpha+(1+\beta(1-2k))(1+\beta(1-k))} \\
&\times\left(\frac{1}{r_0}\right)^{2\beta k/(1+\beta(2k-1))}
 \left(\frac{1}{r_*}\right)^{8\alpha k^2/(1+\beta(1-2k))^2} \,,
\end{split}
\end{equation}
\begin{equation}
d=-\frac{2\beta k}{1+\beta(1-2k)}
-\frac{8\alpha k^2}{1+\beta(1-2k))^2}\,.
\end{equation}
Then, Ricci scalar reads to
\begin{equation}
R=-\frac{(d^2-5d+6)Cr^2r^{-d}-mn(n+1)r^n}{r^2}\,.
\end{equation}
The feature of the Ricci scalar is as follows.
i) To diverge at $r=0$, power terms of $r$ must proportional to
$1/r$. This reads to $0\leq d$ and $n-2\leq0$. This case
our new solution has one physical singular point at origin.
ii) To diverge at $r=\infty$, power terms of $r$ has to be
proportional to $r$. This reads $d\leq 0$ and $0\leq n-2$.
This case our new solution has one physical singular point at
infinity. iii) Both $d=0$ and $n=2$, there is no singular point.
iv) In other case, Ricci scalar will diverge at $r=0$
and $r=\infty$. Then, solution has two physical singular point at
origin and infinity. Higher order curvature invariant such as
$R^{\mu\nu}R_{\mu\nu}$ and $R^{\mu\nu\sigma\lambda}
R_{\mu\nu\sigma\lambda}$ also has same structure.
Thus, our new solution (\ref{eq2-25}) has physical singular point
at $r=0$ or $r=\infty$ that depends on theory to theory.

\begin{figure}[htbp]
\begin{center}
\includegraphics[width=70mm]{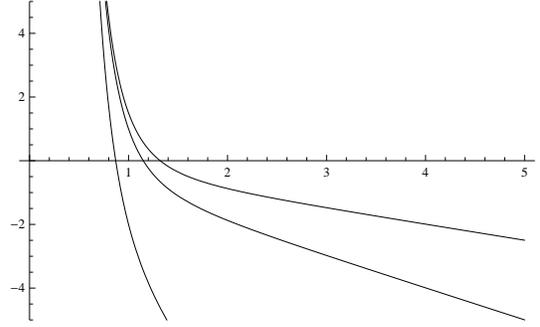}
\end{center}
\caption{The behavior of the metric function when singular point
lies at $r=0$. The parameter sets are $\ell=r_0=r_*=1$ and
$\alpha=1/4~,~\beta=1$ with $k=2$. Left line, middle line and right
line means $m=4,~1,~0.5$ respectively.}
\label{fig4}
\end{figure}

An example of the behavior of the metric function is shown in
Figure~\ref{fig4}. We notice that there is one
horizon in the figure. In the second class solutions, we can
derive the horizon radius as
\begin{equation}
r_H=\left(\frac{m}{C}\right)^{1/(2-d-n)}\,.
\end{equation}
From the figure~\ref{fig4}, we find that the space-time is similar
to the black hole in the de-Sitter space-time. That is why we
conclude that second class solutions can also generate the black
hole. 

%%%%%%%%%%%%%%%%%%%%%%%%%%%%%%%%%%%%%%%%%%%%%%%%%%%%%%%%%%%%%%%%%%
\section{Conclusion and Remarks}
%%%%%%%%%%%%%%%%%%%%%%%%%%%%%%%%%%%%%%%%%%%%%%%%%%%%%%%%%%%%%%%%%%
We derive new solutions of the $2+1$ space-time that coupled to MP
type NLEF with dilaton field. We could derive new circular
symmetric static solutions in general $k$ case. There are two
types solutions that we called ``first class'' and
``second class''. Especially, solutions that belong to second
class, we could get consistent solution at $k=1/2$ theory that 
has no consistent circular symmetric electric
solution without dilaton field. From the singular analyze and
behavior of the metric function, we notice that some parameter
sets will generate black holes.

The future work that we have to do as follows. At first, we have
to investigate more detail structure of the solution. One of them
is thermodynamics of the solution. In the BTZ black hole,
thermodynamical consideration has been performed and the behavior
of thermodynamic quantities are investigated.
The space-time which we derived in our article should be studied
along with the similar thermodynamical context.
Consider the physical structure such as mechanics of our new
solutions seems to interesting. Physical phenomena such as lensing
effect is also interesting. Above analyze will derive physical
viewpoints of our new solutions. 

In addition, to derive more new solutions is important.
We treat electric solutions in our article. Then, we did not derive
magnetic solutions. So considering the magnetic type solution
is meaningful. Second, to consider the rotating effect seems
important. It is known that spinning case has more interesting
property. That is why we have to treat point symmetric type ansatz
in the future work. It seems meaningful to do the same works in
higher dimensional case.  We have to find solutions of higher
dimensional Einstein gravity witch is coupled to non linear
electro-magnetic field with dilaton field.

%%%%%%%%%%%%%%%%%%%%%%%%%%%%%%%%%%
\section*{Acknowledgments}
%%%%%%%%%%%%%%%%%%%%%%%%%%%%%%%%%%%%%%%%%%%%%%%%%%%%%%%%%%%%%%%%%%

%\acknowledgments
%%%%%%%%%%%%%%%%%%%%%%%%%%%%%%%%%%%%%%%%%%%%%%%%%%%%%%%%%%%%%%%%%%%%%%%%%%%
%Acknowledgements
%%%%%%%%%%%%%%%%%%%%%%%%%%%%%%%%%%%%%%%%%%%%%%%%%%%%%%%%%%%%%%%%%%%%%%%%%%%
%\begin{acknowledgments}
I would like to thank K.~Shiraishi for helpful discussions.
I also thank K.~Kobayashi for reading the manuscript.
%\end{acknowledgments}
%%%%%%%%%%%%%%%%%%%%%%%%%%%%%%%%%%%%%%%%%%%%%%%%%%%%%%%%%%%%%%%%%%%%%%

%%%%%%%%%%%%%%%%%%%%%%%%%%%%%%%%%%%%%%%%%%%%%%%%%%%%%%%%%%%%%%%%%%%%%%%%%%%
%thebibliography
%%%%%%%%%%%%%%%%%%%%%%%%%%%%%%%%%%%%%%%%%%%%%%%%%%%%%%%%%%%%%%%%%%%%%%%%%%%

%\bibliographystyle{apsrev}
\bibliographystyle{apsrev4-1}
%\bibliography{}

%%%%%%%%%%%%%%%%%%%%%%%%%%%%%%%%%%%%%%%%%%%%%%%%%%%%%%%%%%%%%%%%%%%%%%
%%%References
%%%%%%%%%%%%%%%%%%%%%%%%%%%%%%%%%%%%%%%%%%%%%%%%%%%%%%%%%%%%%%%%%%%%%%

%%%%%%%%%%%%%%%%%%%%%%%%%%%%%%%%%%%%%%%%%%%%%

\end{document}